\documentclass[aip,rsi,reprint,graphicx]{revtex4-2}

\usepackage{graphicx}
\usepackage{dcolumn}
\usepackage{bm}
\usepackage[utf8]{inputenc}
\usepackage[T1]{fontenc}
\usepackage{mathptmx}
\usepackage{upgreek}
\usepackage{physics}
\usepackage{layouts}
\usepackage{filecontents}
\usepackage[colorlinks,plainpages=false,linkcolor=blue,urlcolor=blue,citecolor=blue,pdfpagemode=UseNone,pdfstartview=FitH]{hyperref}

\begin{document}

\graphicspath{{./Bilder/}}


\title[Quantum stochastic resonance in a single photon emitter]{Quantum stochastic resonance in a single-photon emitter}


\author{H. Mannel$^*$}
\email{hendrik.mannel@uni-due.de}

\author{J. Zöllner$^*$}
\affiliation{Faculty of Physics and CENIDE, University of Duisburg-Essen, 47057 Duisburg, Germany}
\author{E. Kleinherbers}
\affiliation{Department of Physics and Astronomy, University of California, Los Angeles, California 90095, USA}
\author{M. Zöllner}
\affiliation{Faculty of Physics and CENIDE, University of Duisburg-Essen, 47057 Duisburg, Germany}

\author{N. Schwarz}
\affiliation{Faculty of Physics and CENIDE, University of Duisburg-Essen, 47057 Duisburg, Germany}

\author{F. Rimek}
\affiliation{Faculty of Physics and CENIDE, University of Duisburg-Essen, 47057 Duisburg, Germany}

\author{A. D. Wieck}

\author{A. Ludwig}
\affiliation{Lehrstuhl für Angewandte Festk\"orperphysik, Ruhr-Universit\"at Bochum, 44780 Bochum, Germany \\ $^*$ These authors contributed equally to this work.}

\author{A. Lorke}
\author{J. König}

\author{M. Geller}

\affiliation{Faculty of Physics and CENIDE, University of Duisburg-Essen, 47057 Duisburg, Germany}

\date{\today}

\begin{abstract}
Stochastic resonance is a phenomenon in which fluctuations enhance an otherwise weak signal.
It has been found in many different 
systems in paleoclimatology, biology, medicine, and physics.
The classical stochastic resonance due to thermal noise has recently been experimentally extended to the quantum regime, where the fundamental randomness of individual quantum events provides the noise source. 
Here, we demonstrate quantum stochastic resonance in the single-electron tunneling dynamics of a periodically driven single-photon emitter, consisting of a self-assembled quantum dot that is tunnel-coupled to an electron reservoir.
Such highly-controllable quantum emitters are promising candidates for future applications in quantum information technologies.
We monitor the charge dynamics by resonant optical excitation and identify quantum stochastic resonance with the help of full counting statistics of tunneling events in terms of the Fano factor and extend the statistical evaluation to factorial cumulants to gain a deeper understanding of this far-reaching phenomenon. 

\end{abstract}

\maketitle

\section{Introduction}

Noise is usually a nuisance for any measurement as it superimposes and thereby masks the signal one wants to measure.
There are, however, also scenarios in which noise plays an integral role in the system dynamics.
This includes the phenomenon of stochastic resonance\cite{Benzi.1981,Gammaitoni.1998,Wellens.2004}, in which fluctuations enhance an otherwise weak signal.
The effect has been observed in a wide range of different systems, including the periodicity of ice ages \cite{Benzi.1982}, biological signal processing \cite{Chialvo.1993,Douglass.1993,Longtin.1991} and machine learning\cite{Zhai.2023}. 

The paradigmatic model exhibiting stochastic resonance consists of a driven bistable system, in which the energy difference between the two states is modulated in time. 
A potential barrier that is much higher than the driving amplitude prevents the switch between the two states.
Adding fluctuations due to thermal noise allows for thermal activation over the barrier, and the system can switch between the two states to follow the external drive.
The response of the system to the external drive is thus enhanced by the noise.
With increasing noise, the measurement signal becomes more pronounced until, for large fluctuations, the switching of the bistable system becomes independent of the external driving.
This results in a stochastic resonance of the measurement signal, indicated by a large signal-to-noise ratio at the resonance frequency $f_{res}$.

About 30 years ago, a quantum version of the stochastic resonance has been theoretically predicted\cite{Lofstedt.1994,Lofstedt.1994b,Grifoni.1996,Grifoni.1996b}.
It relies on the fact that even in the absence of thermal fluctuations,
stochastic resonance is possible due to quantum noise.
The latter is generated by the inherent phenomenon of quantum tunneling through the potential barrier. The required ingredients  
are quite generic and, therefore, quantum stochastic resonance has been predicted to appear in different systems from general bistable double well potentials\cite{Lofstedt.1994,Lofstedt.1994b,Grifoni.1996,Grifoni.1996b,Makarov.1995,Thorwart.1997,Buchleitner.1998} and specific systems like arrays of superconducting qubits\cite{Huelga.2007}, photoionization of an atom with femtosecond pulses\cite{Singh.2007} and in nuclear magnetic resonance experiments on water\cite{Viola.2000}. However, very few experiments have demonstrated the effect so far. It has been found in electrostatically defined quantum dots\cite{Wagner.2019}, where switching between two charge states is due to electron tunneling between the quantum dot and an electron reservoir.
Quantum stochastic resonance has also been found in a single Fe atom placed on a Cu$_2$N substrate and contacted by a spin-polarized tip of a scanning tunneling microscope\cite{Hanze.2021}. Here, and in both prior works on quantum stochastic resonance, the noise in the system is fixed as it is given by the stochastic process of electron tunneling and the modulation frequency is changed. The Fano factor is used as a measure  of the signal-to-noise ratio.

In this work, we report on the observation of quantum stochastic resonance in a highly controllable single-photon emitter heterostructure consisting of (i) a single self-assembled InAs/GaAs quantum dot (QD) and (ii) an electron reservoir to charge and discharge the dot.
Quantum stochastic resonance is identified as the synchronization of electron tunneling between dot and reservoir driven by periodic modulation of the gate voltage while
monitoring the charge state 
with high-resolution resonance fluorescence\cite{Muller.2007,Melet.2008,Ates.2009,Flagg.2009,Kurzmann.2016b,Hanschke.2020} in real-time\cite{Vamivakas.2010,Kurzmann.2019,Lochner.2020}.
Since single-photon emitters can act as spin-photon interfaces\cite{Yilmaz.2010,Appel.2021,Greve.2012} in future quantum networks\cite{Kimble.2008,Wehner.2018}, our findings could serve as a first step towards regulation of photon streams in continuous-wave laser excitation combined with an external electrical drive. Furthermore we improve the quantitative understanding of quantum stochastic resonance in general. We introduce normalized factorial cumulants of the full counting statistics as a theoretical tool, which allows us to shed light on the non-trivial question of the exact conditions for this quantum-mechanically driven resonance phenomenon. 

\section{Results and Discussion}
\subsection{Charge dynamics in a driven quantum dot}
\subsubsection{Real-time charge detection}

\begin{figure}
	\centering
	\includegraphics{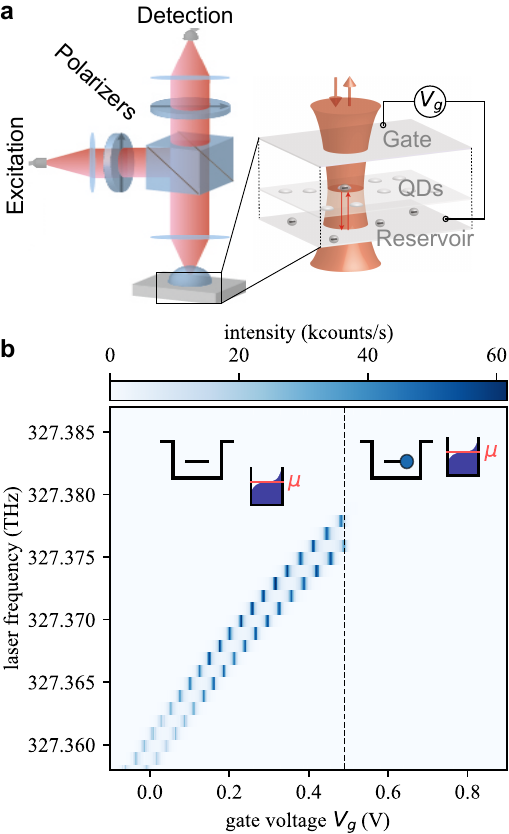}	
	\caption{\label{Fig1} {\bf Resonance fluorescence in a single quantum dot.} \textbf{a} The excitation laser is reflected onto the sample by a 90:10 beamsplitter. Two crossed polarisers are used to suppress the reflected laser light. The QDs are embedded in a p-i-n-diode structure between a gate electrode at the top and an electron reservoir at the bottom of the structure.  A gate voltage $V_g$ can be applied to control the QD charge state and its transition energies. \textbf{b} Exciton transition $X^0$ as a function of the laser excitation frequency and applied gate voltage. At a gate voltage of $V_g=0.5\,$V, the QD is charged with an electron from the reservoir and the $X^0$ transition can no longer be excited by the resonant laser. The insets show schematically the two situations around the transition voltage at $V_g=0.5\,$V, where the chemical potential $\mu$ is below (left inset) and above (right inset) of the lowest quantum dot state.}
\end{figure}

A schematic illustration of our sample structure and experimental setup is shown in Fig.~\ref{Fig1}a. A self-assembled InAs/GaAs dot is embedded in a p-i-n diode structure for electrical control (details in methods section and in supplementary note 1) and cooled in a bath cryostat to 4.2 K\cite{Lochner.2019}.
The dot is charged and discharged by tunneling of electrons from or to an electron reservoir, i.e., a highly doped GaAs layer below the dot. A gate voltage $V_{g}$ allows us to tune the occupation probability of the QD from zero to one, as the lowest electron state is energetically shifted through the Fermi edge of the electron reservoir (see upper insets in Fig.~\ref{Fig1}b).
The charge state is detected in real-time by means of resonance fluorescence from the neutral exciton transition\cite{Kurzmann.2019}. 

We use the dark-field technique (crossed excitation-collection polarization) to separate the reflected laser light from the QD signal\cite{Vamivakas.2009,Houel.2012,Matthiesen.2013,Kuhlmann.2013}. The light emitted from the QD is detected by an avalanche photo diode (APD) that serves as a single-photon detector (see methods section). The detected photons are accumulated
over a binning time $t_{bin}$ to obtain a measure of the fluorescence brightness\cite{Kerski.2023}. 

Figure \ref{Fig1}b shows the resonance fluorescence (RF) of the exciton transition ($X^0$) with its fine-structure splitting\cite{Gammon.1996,Bayer.2002} as a function of gate voltage $V_g$ and laser frequency. 
Due to the quantum confined-Stark effect, the applied gate voltage shifts the frequency of the exciton resonance towards higher energies as the gate voltage is increased. 
In addition, the energy of the quantum dot states will decrease with respect to the chemical potential $\mu$ of the electron reservoir until, around $V_g=0.5\,$V, an electron can tunnel into the s-state of the dot, marked by the dashed line in Fig.~\ref{Fig1}b and schematically illustrated by the inset in the upper right corner. 
This shifts the optical resonance, and the neutral exciton transition vanishes. 
Close to this gate voltage, the QD charge fluctuates in time and the exciton transition can be used as a sensitive optical detector to observe all tunneling events in real-time. 

\begin{figure}
	\centering
	\includegraphics{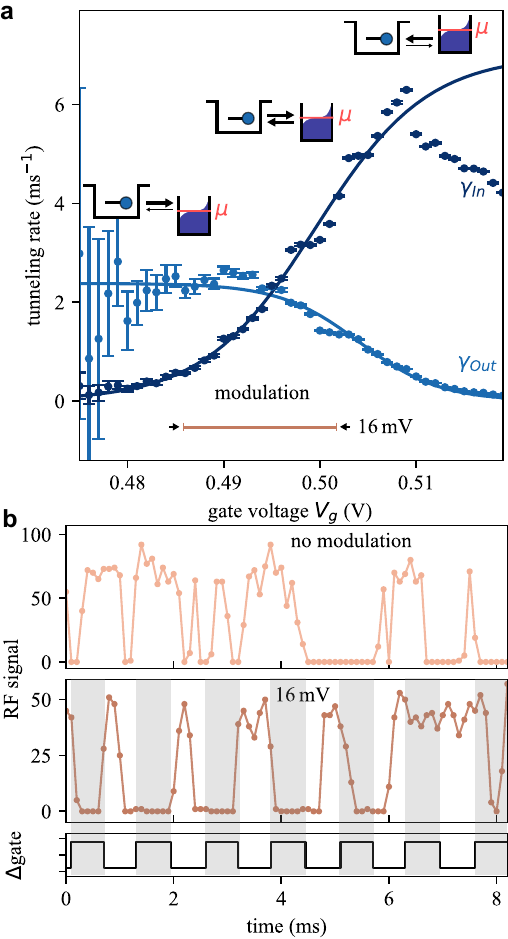}
	\caption{\label{Fig2} {\bf The electron dynamics, revealed in the optical resonance fluorescence signal.} {\bf a} Electron tunneling rates $\gamma_{In}$ and $\gamma_{Out}$ as a function of the gate voltage $V_g$. Solid lines are fits to the data, given by the Fermi-distribution in the electron reservoir. The insets illustrate the continuous increase of $\gamma_{In}$ with increasing gate voltage, while $\gamma_{Out}$ decreases with increasing chemical potential with respect to the QD level. Around the gate voltage where both rates are equal, the voltage modulation for the quantum stochastic resonance takes place. The error bars are derived via $R^2$ of the exponential fit presented in Ref. \onlinecite{Zollner.2024}. {\bf b} Telegraph signals with a binning of $100\,\mu$s, showing the single electron tunneling events between the quantum dot and the reservoir as a switching on and off of the resonance fluorescence signal. For no modulation the electron tunnels in and out randomly with fixed rates $\gamma_{In}$ and $\gamma_{Out}$. When the modulation of the gate voltage is on, correlations start to occur between the electron tunneling and $V_g$. This is shown with shaded areas as a guide to the eye where a switching event should occur for deterministic transport. The amplitude of the a.~c.~drive is indicated as a scale bar in {\bf a}.}
\end{figure}

The gate voltage range between the QD being uncharged and charged is given by the reservoir temperature of $4.2\,$K and a gate-voltage-to-energy conversion factor ('lever arm') of $15.2 \,\text{mV}\text{meV}^{-1}$. 
This is demonstrated in Fig.~\ref{Fig2}a, which displays the tunneling rates $\gamma_{In}$ and $\gamma_{Out}$ for tunneling into and out of the QD, that were determined by means of the pulsed measurement scheme presented in Ref.~\onlinecite{Kurzmann.2016b,Zollner.2024}.
For $V_g < 0.47\,$V, $\gamma_{In}$ vanishes and the QD is mostly empty, while for $V_g > 0.52\,$V, $\gamma_{Out}$ vanishes, and the QD becomes occupied with a single electron.
In between, the average QD occupation, given by $\gamma_{In} / (\gamma_{In}+\gamma_{Out})$, changes continuously from $0$ to $1$ with increasing gate voltage.
The energetic position of the QD state with respect to the chemical potential $\mu$ of the electron reservoir is shown in three schematic insets. At $V_g=494\,$mV, the tunneling rates $\gamma_{In}$ and $\gamma_{Out}$ are approximately equal, with electrons randomly tunneling back and forth between the QD and reservoir. Note that here no Auger process is possible because only the exciton (and not the trion) transition is driven.
A charged QD yields a low ("off") fluorescence signal while the uncharged dot leads to a bright ("on") fluorescence.
The result is a random telegraph signal in the optical response of the QD\cite{Kurzmann.2019}.

To drive the system, we modulate the gate voltage with a square function, as shown at the bottom of Fig.~\ref{Fig2}b and also indicated by the modulation scale bars in Fig.~\ref{Fig2}a.
Stochastic resonance can be qualitatively characterized by the switching behavior of the charge state: the switching becomes more regular at resonance while it is dominated by fluctuations away from it.

In Fig.~\ref{Fig2}b, a photon stream without modulation is shown, where the electron tunnels in and out randomly.
With a modulation amplitude of $16\,$mV, though, even the unaided eye can discern that the low/high value of the gate voltage (white/grey region in Fig.~\ref{Fig2}b) is predominantly accompanied with a large/small RF signal in the telegraph stream.
Hence, the charge switching is synchronized with the driving voltage, indicating stochastic resonance.

\subsubsection{Full counting statistics}

In Fig.~\ref{Fig2}b, center, the synchronization between voltage drive and charge-state switching is evident.
A systematic analysis of quantum stochastic resonance, however, requires a quantitative tool not only to characterize how regular the switching occurs but also to identify the resonance condition, i.e. to find (for given gate voltage values) the modulation frequency $f_{mod}$ for which the switching is most regular. 
For this, a statistical analysis of the switching behavior needs to be performed.

\begin{figure}
	\centering
	\includegraphics{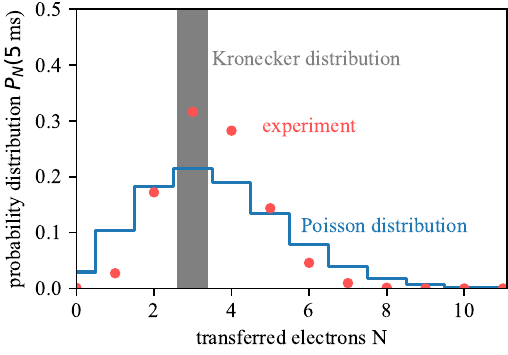}
	\caption{\label{Fig3} 
    {\bf Probability distribution of electron transport.} 
    Probability distribution $P_N(\Delta t=5\,$ms)
    (red dots) for the number of electrons transferred at a modulation frequency of $f_{mod}=796\,$Hz with amplitude $16\,$mV. The experimental data is compared to a Kronecker
    distribution (gray), corresponding to completely regular switching, and a Poisson distribution (blue line) as an completely uncorrelated transport. The measured distribution is narrower than a Poisson distribution for the same mean value. The reduced width of a distribution indicates a more regular electron transport.
    }
\end{figure}

To address the switching dynamics, we make use of the framework of full counting statistics, applied to the number of switching events within a given time span.
The central quantity is the probability distribution $P_N(\Delta t)$ that $N$ switching events have taken place within an interval of length $\Delta t$.
To be specific, we decide to count only tunneling-out events.
The alternative choices of counting only tunneling-in events or counting both types of tunneling events (as it was done in Ref.~\onlinecite{Wagner.2019}) would render our conclusions about the stochastic resonance unchanged.
An example for the distribution $P_N(5 \,\text{ms})$ for a modulation amplitude of $16\,$mV and frequency of $f_{mod}=796\,$Hz is shown in Fig \ref{Fig3}.
The distribution $P_N(5 \,\text{ms})$ (red dots) is broadened as compared to a completely regular behavior of one tunneling-out event per modulation cycle (gray).
This reflects the stochastic nature of tunneling.
The measured distribution is, however, much sharper than a Poisson distribution with the same mean value (blue line).
The latter would be expected if all tunneling-out events were independent of each other.
The reduced width of the measured distribution, thus, indicates that the switching behavior has become more regular. We remark that this is not yet a unique indicator of stochastic resonance, since even in the undriven system correlations (due to Pauli exclusion and electron-electron interactions) lead to sub-Poissonian statistics.

A more quantitative comparison of a given distribution, within the completely random and completely regular limits, also used by previous experimental studies\cite{Wagner.2019,Hanze.2021}, is provided by the signal-to-noise ratio or its inverse, the Fano factor
\begin{align}
    F = \frac{\langle N^2 \rangle - \langle N \rangle^2}{\langle N \rangle} \, ,
    \label{eq:Fano}
\end{align}
i.e. the ratio of the variance $\sigma^2=\langle N^2 \rangle - \langle N \rangle^2$ and the mean value $\lambda = \langle N \rangle$ for the number $N$ of measured switching events.
If the tunneling-out events occurred completely independent of each other, the full counting statistics would be Poissonian, which implies a Fano factor of $1$.
However, correlations are always present since after a tunneling-out event an electron has to tunnel in before the next tuneling-out can occur.
For the undriven case, the Fano factor is known\cite{Hershfield.1993,Blanter.2000,Bagrets.2003,Gustavsson.2006, Garreis.2023} to be $F=\frac{\gamma_{In}^2+\gamma_{out}^2}{(\gamma_{In} + \gamma_{Out})^2}$, which ranges between values of $F=1$ for $\gamma_{In}\ll \gamma_{Out}$ or $\gamma_{Out}\ll \gamma_{In}$ and $F=0.5$ for $\gamma_{In}=\gamma_{Out}$.
The measured value of $F=0.35$ in Fig.~\ref{Fig4}, therefore, indicates a more regular switching behavior in the presence of driving.
In the limit of completely regular switching, the Fano factor would be $F=0$.
With this tool at hand, stochastic resonance can be defined by the minimum of $F$ as a function of the driving frequency for fixed switching rates or as a function of switching rates for fixed driving frequency\cite{Wagner.2019}.

The characterization of such a complex process as quantum stochastic resonance, using only two numbers $\lambda$ and $\sigma^2$, however, seems to be limited. 
Deeper insight is expected from a more complete theoretical evaluation.
As one possibility, the spectral properties of the noise, expressed in terms of a frequency-dependent Fano factor, has been employed for the analysis of the quantum stochastic resonance\cite{Hussein.2020}.
Here, we suggest an alternative route that is based on the idea that a distribution function can be completely characterized in terms of cumulants.
The mean value $\lambda$ and the variance $\sigma^2$ are nothing but the first- and second-order ordinary cumulants.
They are sufficient to fully describe a Gauss distribution.
Deviations from a Gauss distribution are expressed in terms of third- and higher-order ordinary cumulants.
They have been measured in quantum-dot systems\cite{Gustavsson.2007,Flindt.2009,Komijani.2013} up to 20-th order.
Our aim, however, is to measure deviations from a Poisson rather than a Gauss distribution.
Therefore, we make use of factorial cumulants\cite{Beenakker.2001,Johnson.2005,Kambly.2011,Stegmann.2015,Stegmann.2016,Brange.2019}, which are better suited for discrete rather than continuous stochastic variables.
They prevent unwanted universal oscillations\cite{Flindt.2009,Komijani.2013} and are resilient against detector imperfections\cite{Kleinherbers.2022}.
They have proven successful in determining the rates for spin relaxation\cite{Kurzmann.2019} as well as Auger recombination and spin-flip Raman scattering\cite{Kleinherbers.2023} from random telegraph signals such as the ones investigated here.

The factorial cumulants $C_{F,m}$ of order $m$ can be calculated using the generating function
\begin{align}
	S_F(z)=\ln \sum_{N}z^N P_N,
\end{align}
i.e., the logarithm of the $z$-transform of the probability distribution $P_N$, and taking their derivatives with respect to the counting variable $z$,
\begin{align}
	C_{F,m}=\left.\left[\partial^m_z S_F(z)\right]\right|_{z=1} \, .
    \label{eq:factorial_cumulants}
\end{align}
Explicit expressions for the first four factorial cumulants in terms of moments $\langle N^m \rangle$ are given in Eqs.~(S2a-d), see supplementary note 5.

Using the factorial cumulants, the Fano factor can be alternatively expressed as 
\begin{align}
	F =1+\frac{C_{F,2}}{C_{F,1}} \, ,
    \label{eq:Fano factor}
\end{align}
where the first-order factorial cumulant $C_{F,1}=\langle N \rangle = \lambda$ is the mean value of counts.
Higher-order factorial cumulants vanish for a Poisson distribution of completely uncorrelated discrete events, $C_{F,m}^{Poisson} = 0$ for $m \ge 2$.
In the opposite limit of fully regular switching behavior, the distribution is described by a Kronecker delta, $P_N^{\delta}=\delta_{N,\lambda}$ which yields $C_{F,m}^{\delta}=(-1)^{(m-1)}(m-1)!\lambda$.

For a quantitative measure of the regularity of the switching behavior, it is useful to normalize the measured factorial cumulants $C_{F,m}$ with $C_{F,m}^{\delta}$.
We, thus, define for $m\ge 2$ the normalized factorial cumulants
\begin{align}
    x_m := \frac{(-1)^{m-1}C_{F,m}}{(m-1)!C_{F,1}} \, ,
\end{align}
that are all $0$ for a Poisson and $1$ for a Kronecker distribution.

To infer the factorial cumulants (and Fano factor) from the measured random telegraph stream of 15 min for a given time interval length $\Delta t$, we first determine the probability distribution $P_N(\Delta t)$ for the number $N$ of tunneling events.
Then, we calculate the moments $\langle N^m \rangle$ up to $m=4$ and insert them into Eq.~(\ref{eq:Fano}) for the Fano factor and Eqs.~(S2a-d) shown in supplementary note 5 for the factorial cumulants, respectively.

\subsection{Quantum stochastic resonance}

Having established that normalized factorial cumulants 
$x_m$ provide a quantitative measure how regular or random a process is, we define the quantum stochastic resonance frequency as the frequency that maximizes $x_m$. We will show below that the frequency that maximises $x_m$ depends on the order $m$ of the chosen factorial cumulant, raising the question of how rigorous a resonance frequency can be defined in the context of stochastic resonance.

\subsubsection{Fano factor}

We begin with a discussion of $x_2$.
It is related to the Fano factor by $F=1-x_2$, i.e. a maximum in $x_2$ is equivalent to a minimum in $F$.
In the field of quantum optics, $-x_2=C_{F,2}/C_{F,1}=:Q $ is also known as the Mandel $Q$-parameter to characterize deviations from Poissonian photon statistics~\cite{Mandel.1979}.

The distribution $P_N(\Delta t)$ depends on the length $\Delta t$ of the time interval in which the tunneling-out events are counted.
Furthermore, the accuracy with which the random telegraph signal reflects the actual charge switching of the system depends on the binning time $t_{bin}$, which is the time interval used to group the photon arrival events. 
The binning time $t_{bin}$ is chosen large enough to distinguish the charged from the uncharged state by a clear separation of the two peaks in the histogram of the RF counts per bin, see Fig.~S2 in supplementary note 2.
Both times, $\Delta t$
and $t_{bin}$, can be chosen a posteriori from the measured stream of individual photon counts\cite{Kerski.2023}.
They introduce two new time scales that are unrelated to stochastic resonance.
We must therefore optimize them in our quest for stochastic resonance to avoid artifacts.

\begin{figure}
	\centering
	\includegraphics{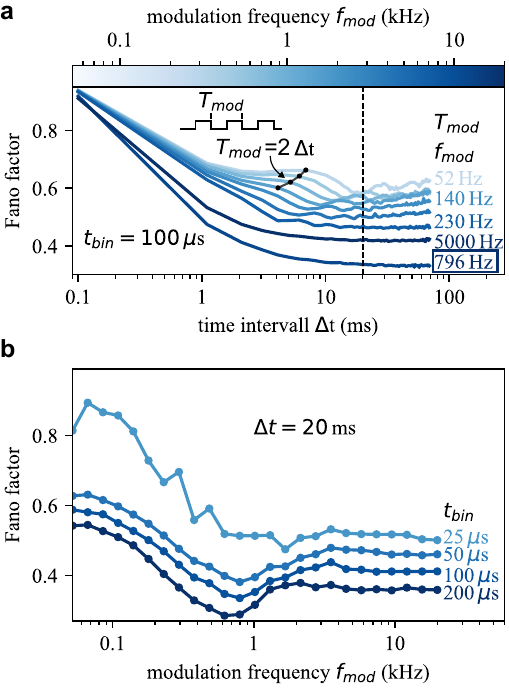}	
	\caption{\label{Fig4} {\bf Time evolution of the Fano factor.} \textbf{a} The Fano factor versus the time interval $\Delta t$ for selected modulation frequencies from $52\,$Hz (light blue) to $5\,$kHz (dark blue) and a modulation amplitude of $16\,$mV. At a driving frequency of 796Hz, the Fano factor is at a minimum for all $\Delta t$, indicating stochastic resonance around this frequency. The dashed line is at the time intervall of $\Delta t=20\,$ms, where we define the long time limit (see subsection Fano factor). The maxima of the Fano factor marked with black dots are artefacts that occur when the time interval $\Delta t$ is half of the modulation period $T_{mod}$; schematically shown in the inset and explained in more detail in the main text. \textbf{b} The long time limit of the Fano factor for different binning times $t_{bin}$ of the single photon stream. The decrease in the Fano factor at $\approx800\,$Hz is due to the quantum stochastic resonance. 
	}
\end{figure}

In Fig.~\ref{Fig4}a, we show the Fano factor as a function of $\Delta t$ for different modulation frequencies ranging from $52$ Hz up to $5$ kHz.
For all data sets that each comprise a time span of $15\,$min, we used a modulation amplitude of $16\,$mV and a binning time of $t_{bin}=100\, \mu$s. Each data set contains 600k events on average.
For very short intervals $\Delta t$, most intervals accommodate no or, with a small probability, only one tunneling-out event.
This can be described by a Bernoulli distribution $P_N^{Bernoulli}=(1-q)\delta_{N,0} + q \delta_{N,1}$ with $0\le q \ll 1$, for which the Fano factor $F=1-q$ approaches the Poissonian value $F\to 1$ for $\Delta t \to 0$ (which implies $q \to 0$). 
As a result, operating in the short-time limit is inappropriate for unraveling quantum stochastic resonance.
Instead, it is advantageous to go to the long-time limit, in which each interval contains many counts of tunneling-out events such that the Fano factor becomes independent of $\Delta t$.
In the following, we choose $\Delta t = 20\,$ms (indicated by the vertical dashed line in Fig.~\ref{Fig4}a) to ensure that, on the one hand, the long-time limit is already reached and that, on the other hand, slow, long-term fluctuation in the experimental setup will not become relevant yet.

Since the period $T_{mod} = 1/f_{mod}$ of the gate modulation already introduces a time scale, one may be tempted to choose this time scale for $\Delta t$ as well.
This, however, may introduce artifacts as shown in Fig.~\ref{Fig4}a, where the choice $\Delta t = T_{mod}/2$ is indicated by black dots.
We find that, at least for low frequencies, the Fano factor is enhanced around this choice of $\Delta t$. To identify quantum stochastic resonance, we vary the modulation frequency $f_{mod}$.
Fig.~\ref{Fig4}a shows that for all values of $\Delta t$ the Fano factor first decreases with increasing modulation frequency (from $f_{mod}=52\,$Hz to $f_{mod}=796\,$Hz), and subsequently increases when the modulation frequency is set to $5\,$kHz (or larger).
This is shown more clearly in Fig.~\ref{Fig4}b, which depicts the Fano factor as a function of the modulation frequency $f_{mod}$ for different values of the binning time $t_{bin}$. The interval length is fixed at the long-time limit of $\Delta t = 20\,$ms.
For a binning time of $t_{bin}=100\, \mu$s, as chosen in Fig.~\ref{Fig4}a, the Fano factor has a minimum at $f_{mod} = 796\,$Hz.

The binning time $t_{bin}$ affects the accuracy of the charge-state switching detection. A short binning time provides high time resolution, however, increases noise, leading to false event detection. This explains why at $t_{bin}=25 \, \mu$s, the Fano factor shows no clear minimum. Increasing $t_{bin}$ reduces false detections. However, this comes at the expense of time resolution, which results in fast events being missed. At $t_{bin} = 200 \, \mu$s, the Fano factor shows a clear minimum, however, slightly shifted. As a compromise, we choose $t_{bin} = 100 \, \mu$s, the smallest value where the resonance remains well pronounced.

\begin{figure*}
	\centering
	\includegraphics{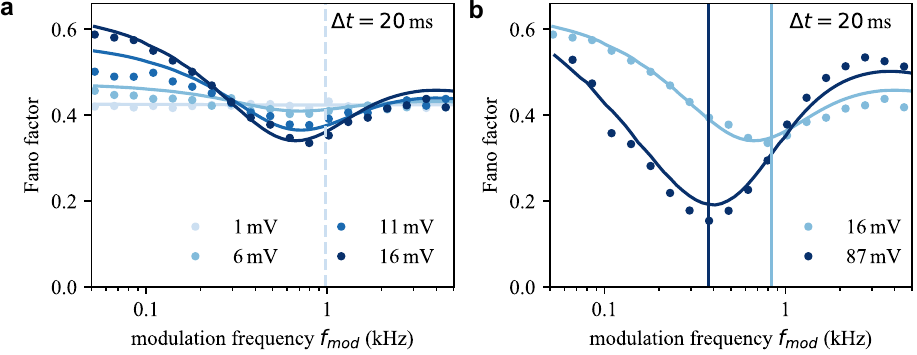}	
	\caption{\label{Fig5} \textbf{Quantum stochastic resonance indicated by the Fano factor.}
    \textbf{a} Fano factor as a function of the modulation frequency for different modulation amplitudes (data points) compared with theoretical model curves (solid blue lines). The model has no fitting parameters. The modulation is symmetrical around $V_g=496\,$mV, where $\gamma_{In}=\gamma_{Out}$. The minimum around $\approx800\,$Hz
    is due to 
    quantum stochastic resonance. The dashed vertical line is given by Eq.~\ref{eq:f_res_linear} for a $1\,$mV modulation. \textbf{b} The Fano factor as a function of the modulation frequency for $16\,$mV and $87\,$mV. The exact values of the upper, lower and average value of the gate voltage are given in Table S1 in supplementary note 3. 
    The vertical lines are given by Eq.~\ref{eq:f_res_nonlinear} for a $16\,$mV and $87\,$mV modulation. 
	}
\end{figure*}

In Fig.~\ref{Fig5}a and b, we show the Fano factor as a function of the modulation frequency for five
different modulation amplitudes. 
In all cases, the average gate voltage is chosen close to the value where the in- and out-tunneling rates are equal to each other (the exact values of the upper, lower and average value of the gate voltage is given in Table S1 in supplementary note 3).
A modulation amplitude of $1\,$mV is not sufficient to identify stochastic resonance.
With increasing amplitude, however, a pronounced minimum develops around $f_{mod} \approx 800\,$Hz. For a modulation amplitude of $16\,$mV, the Fano factor drops below 0.35, see Fig.~\ref{Fig5}a. 

In Fig.~\ref{Fig5}b, we compare the Fano-factor for a moderate modulation amplitude (16 mV) to that for a high amplitude (87 mV). For the high amplitude, we not only observe  a deep minimum, down to $F = 0.15$.
We also find that the position of the stochastic resonance is strongly shifted towards a lower frequency of about 500 Hz, which is discussed in the next section.
We note an experimental detail here: Due to the larger modulation amplitude, the laser is no longer in resonance with the exciton $X^0$ transition (see Fig.~\ref{Fig2}a) during the half cycle with the larger gate voltage.
As a result, the electron dynamics by photon counting is only detected during the lower gate voltage phase.

For all voltage amplitudes, we compare the measured data (dots) with a theoretical model (lines) that takes the finite time resolution of the detector into account, see Methods.
We find that both the position and the depth of the minimum are well reproduced, see Fig.~\ref{Fig5}.
Missed events due to the finite time resolution of the detector reduce the measured $C_{F,1}$ while $C_{F,2}$ remains almost unaffected due to the resilience of the higher-order factorial cumulants against detector imperfections~\cite{Kleinherbers.2022}.
This decreases the Fano factor but hardly moves the position of the minimum.

\subsubsection{Resonance frequency}

The value of the resonance frequency $f_{res}=1/T_{res}$ depends in a complicated way on all four rates $\gamma_{In,low}$, $\gamma_{In,high}$, $\gamma_{Out,low}$ and $\gamma_{Out,high}$, where $\gamma_{In,low/high}$ is the rate for charging the quantum dot during the low/high value of the gate voltage and, similarly $\gamma_{Out,low/high}$, the rate for discharging.
There are, however, two limits in which simple analytical expressions can be derived (shown in Fig.~\ref{Fig5} as vertical lines).
To achieve this, we compare the modulation period $T_{mod}=1/f_{mod}$ with the waiting times for the quantum dot being in a given state before it switches, i.e., the inverse of the corresponding rate.

For small modulation amplitudes, i.e. in the linear-response regime, we find the condition for stochastic resonance by requiring that there is exactly one tunneling-out event (and, therefore, also one tunneling-in event) per modulation cycle.
For this, the modulation period $T_{mod}$ has to be equal to the sum of the waiting times of the quantum dot to be charged and discharged, respectively, where
we can use the tunneling rates $\gamma_{In}$ and $\gamma_{Out}$ of the undriven system and find\cite{Talkner.2003} 
\begin{align}
\label{eq:f_res_linear}
    T_{res} = \frac{1}{\gamma_{In}} + \frac{1}{\gamma_{Out}}
\end{align}
since $\gamma^{-1}_{In}$ is the waiting time for the QD being uncharged and $\gamma^{-1}_{Out}$ for being charged.
For the present experiment with $\gamma_{In} = \gamma_{Out} =: \gamma$, this reduces to the condition $2f_{res} = \gamma$.\cite{Talkner.2005}
The value of this estimate for a modulation amplitude of $1$ mV, where stochastic resonance is not yet visible, is shown as a dashed line in Fig.~\ref{Fig5}a.
Already for $6$ mV modulation amplitude, the linear-response estimate can no longer be used since the low- and high-voltage tunneling rates strongly differ from each other (and therefore also from the undriven case).

A simple analytic expression can also be found in the regime $\gamma_{In,low} \ll \gamma_{Out,low}$ and $\gamma_{Out,high} \ll \gamma_{In,high}$, for which a large modulation amplitude is necessary (but not sufficient).
In the extreme limit of $\gamma_{In,low}=0$ and $\gamma_{Out,high}=0$, a perfectly regular switching behavior with exactly one discharging event per cycle is achieved in the limit $T_{mod}\to \infty$.
This implies (for $\Delta t = T_{mod}$) a vanishing Fano factor in the limit $f_{mod}\to 0$ and, thus, qualitatively explains why the minimum of the Fano factor is shifted towards lower frequencies for large modulation amplitudes.
For small but finite $\gamma_{In,low}$ and $\gamma_{Out,high}$, the regular switching behavior is destroyed once the quantum dot can be charged during the low-voltage ($T_{mod}/2 = \gamma_{In,low}^{-1}$) or discharged during the high-voltage phase ($T_{mod}/2 = \gamma_{Out,high}^{-1}$).
Taking the smaller of the two modulation periods and reducing the modulation period by a factor $1/2$ to ensure that the destruction of the regular switching behavior has not yet set in, we arrive at
\begin{align}
\label{eq:f_res_nonlinear}
    T_{res} = \frac{1}{\max \{\gamma_{In,low}; \gamma_{Out,high} \} } \, ,
\end{align}
i.e. the larger of the two small rates determines the resonance condition.
The vertical lines in Fig.~\ref{Fig5}b demonstrate a good agreement of the estimate with the measured data.

\subsection{Higher-order factorial cumulants}

\begin{figure}
	\centering
	\includegraphics{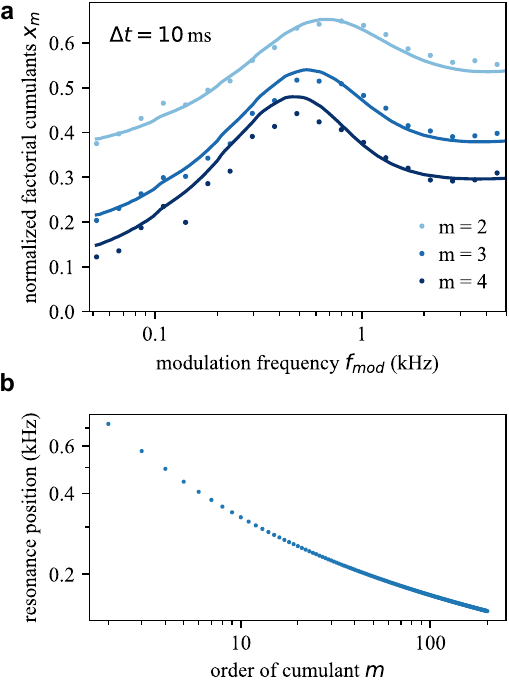}	
	\caption{\label{Fig6} \textbf{Quantum stochastic resonance indicated by factorial cumulants.} 
    \textbf{a} The 
    second, third and fourth normalized factorial cumulant $x_m$  
    as a function of the modulation frequency for the experimental data as blue dots and the  model as solid blue lines at a modulation amplitude of $16\,$mV and a time interval length of $\Delta t=10$ms. 
    A value of $x_m=1$ would indicate completely regular and $x_m=0$ completely random switching, respectively. 
    A resonance can be seen at around $800\,$Hz,
    shifting towards lower modulation frequencies with higher order of the cumulant. \textbf{b} The modeled resonance frequency $f_{res}$ as a function of the cumulant order for an ideal detector. Note how $f_{res}$ decreases towards zero with increasing order and no longer satisfies the simple relation $2f_{res} = \gamma$ (see Eq.~\ref{eq:f_res_linear}). 
    }
\end{figure}

Going beyond the second-order normalized factorial cumulant $x_2$ (or equivalently the Fano factor or the Mandel $Q$-parameter),
we now also consider the higher-order terms $x_3$ and $x_4$ as a function of the modulation frequency $f_{mod}$.
This is shown in Fig.~\ref{Fig6}a for $16\,$mV modulation amplitude and a time interval length of $\Delta t=10\,$ms.
We find good agreement between experimental data (dots) and theoretical model (lines) without any fitting parameter.
In each order $m$, we find a clear resonance.
This nicely shows that the influence of stochastic resonance affects the statistical distribution far beyond the second moment or the Fano factor.

We find that the resonance frequency decreases with increasing order $m$.
While experimentally we can only access the first few orders (Fig.~\ref{Fig6}a), we can calculate them within our theoretical model up to very high order, see Fig.~\ref{Fig6}b, and find a pronounced shift with no obvious lower bound.
A physical understanding of this intriguing behavior remains an open question. It demonstrates, however, that the intuitive assessment that stochastic resonance occurs at the frequency given by the switching rate is incomplete. 

Finally, we remark that higher-order ordinary cumulants are not useful for identifying stochastic resonance. 
They vanish ($C_m=0$ for $m\ge 2$) for a Kronecker distribution ($P_N^{\delta}=\delta_{N,\lambda}$), but these zeroes of $C_m$ as a function of the frequency are masked by trivial zeroes that arise due to universal oscillations~\cite{Flindt.2009}, see supplementary note 9.

\section{Conclusions}

In conclusion, we have demonstrated quantum stochastic resonance in the dynamics of single-electron tunneling of a periodically driven two-state quantum system which consists of an electron reservoir and a quantized state in a single quantum emitter. 
The tunneling events are detected in real-time using resonance fluorescence on the exciton transition as an optical readout scheme. 
The telegraph signal of the tunneling events allows us to analyze the quantum stochastic resonance using full counting statistics.
Quantum stochastic resonance is indicated by a dip in the Fano factor as well as a peak in the normalized factorial cumulants as a function of the driving frequency.
We discuss the dependence of the resonance frequency on the modulation amplitude and find simple expressions both for linear response and the limit of strong nonequilibrium.
Furthermore, we find a shift of the resonance frequency as a function of the order of the factorial cumulants.
This shows that the question about the exact condition for stochastic resonance is non-trivial, and we expect that our study will trigger further investigations about the origin as well as the qualitative and quantitative description of quantum stochastic resonance.
We advocate to apply normalized factorial cumulants for a quantitative analysis of quantum stochastic resonance in other devices as well.

\section*{Methods}
\subsection*{QD sample structure}
The sample is grown by molecular beam epitaxy and contains a single layer of self-assembled InAs/GaAs quantum dots (QDs) embedded within a GaAs matrix. The QDs are incorporated into a p-i-n-diode structure for precise charge state control (see supplementary note 1). The n-doping of the diode serves as a three-dimensional electron reservoir (back gate), allowing controlled charging of the QD or setting it to a regime where an electron can tunnel in and out of the QD by adjusting the gate voltage $V_g$ and the shift of QD exciton energy levels through the quantum-confined Stark effect. To enhance collection efficiency within the spectral range, the sample includes a Bragg reflector of GaAs/AlGaAs layers. Moreover, the photon collection efficiency is improved by a zirconia solid immersion lens on top  of the sample. 
\subsection*{Resonance fluorescence measurement technique}
We use resonance fluorescence to optically detect the charge state of the QD, which is achieved using a confocal microscopy setup with a 90:10 beam splitter head and a sample mounted in a helium bath cryostat at 4.2 K. The resonant excitation and the detection of the scattered light from the laser and QD pass along the same path through a 0.68 NA objective lens. To suppress the scattered laser light from the QD photons, we use the standard cross-polarization technique, where two linear polarizers before and after the beam-splitter suppress the laser light by at least a factor of $10^7$.  The photons are detected by a single photon detector (an avalanche photo diode), and we achieve a laser-background-to-QD-signal ratio well below 1\%. With the particular QD investigated here, a maximum photon count rate of $4\,$MCounts/s ($2\,$MCounts/s per fine-structure splitted exciton transition) was achieved.
\subsection*{Post processing of the photon stream}
Single photons are detected using an APD and time-correlation with a time-to-digital converter, generating a list of arrival times for each photon-counting event. The telegraph signal of the electron tunneling events is derived by counting the number of photon events within a selected binning time $t_{bin}$. The time resolution of the photon stream can be adjusted after the actual measurement. Post processing enables us to find an optimal binning time that achieves the highest possible bandwidth while still distinguishing the excitonic on- and off-state of the QD\cite{Kerski.2023} (see supplementary note 2).
\subsection*{Theoretical Model}
To calculate the factorial cumulants, we first introduce the state-resolved probability distribution $P_{\chi,N}(t)$ for the two charge states $\chi =0, 1$, defined as the combined probability that at time $t$ the quantum dot is in charge state $\chi$ and $N$ tunneling-out events have occurred since the beginning of counting at time $t_0$.
Their $z$-transforms $P_{\chi,z}:=\sum_N z^N P_{\chi,N}$ are governed by the master equation~\cite{Kambly.2011,Stegmann.2015,Kleinherbers.2022}
\begin{align}
\begin{split}
    \dot P_{0,z}(t) &= -\gamma_{In}(t)P_{0,z}(t)+z \gamma_{Out}(t)P_{1,z}(t)\\
	\dot P_{1,z}(t) &= \gamma_{In}(t)P_{0,z}(t)-\gamma_{Out}(t)P_{1,z}(t)
\end{split}
\end{align}
with rates $\gamma_{In}(t)$ and $\gamma_{Out}(t)$ for tunneling in and tunneling out, respectively. 
The factor $z$ in the second term of the upper equation is referred to as the counting variable.
We solve these differential equations for a square-function time dependence of the rates.
To account for the limited time resolution of the detector, we discretize the differential equations with time steps of length $t_{bin}$ according to Ref.~\onlinecite{Kleinherbers.2022}. 
The sum $P_z(t):=P_{0,z}(t)+P_{1,z}(t)$ of the results $P_{0,z}(t)$ and $P_{1,z}(t)$ is the $z$-transform of the distribution $P_N(t)$, which yields the generating function 
\begin{align}
    S_F(z,t)=\ln P_z (t) \, .
    \label{eq:generating_function_solution}
\end{align}
To calculate the factorial cumulants for an interval of lengths $\Delta t$, we set $t=t_0+\Delta t$, average the distribution $P_z$ for fixed $\Delta t$ over the initial times $t_0$, take the logarithm and, then, apply Eq.~(\ref{eq:factorial_cumulants}).
A more detailed description of the theoretical modeling is given in supplementary note 7. 

\vspace{.5cm}
\section*{Author contributions}
H.M., M.Z., N.S., F.R. carried out the experiments on resonance fluorescence.
J.Z., E.K. and J.K. developed the theory and analysed the data by full counting statistics.
H.M., A.Lo and M.G. designed the experiments and analysed the data.
A.Lu. and A.D.W designed and grew the sample.
J.Z., H.M., A.Lo., J.K., M.G. wrote the manuscript. All authors contributed to the discussion of the results and the preparation of the manuscript.

\section*{Competing interests}

The authors declare no competing interests.

\section*{Data Availability}
The data that support the findings of this study are available from the corresponding author upon reasonable request.
 
\begin{acknowledgments}
	This work was funded by the Deutsche Forschungsgemeinschaft (DFG, German Research Foundation) through SFB 1242 with Project-ID 278162697 (A.~L., J.~K. and M.~G.), and the individual research grant 383065199 (M.~G.~and A.~Lu.). The Mercator Research Center Ruhr (MERCUR) is gratefully acknowledged for support within the project No.~Ko-2022-0013 (A.~L., J.~K. and M.~G.).  A. Lu. and A. D. W. acknowledge support by DFG-TRR160, BMBF - QR.X KIS6QK4001, and the DFH/UFA CDFA-05-06.
\end{acknowledgments}

\section*{References}

\bibliographystyle{naturemag}
\bibliography{StochRes-PaperBib-Paul}

\end{document}


\title{Supplementary information: 
\\Quantum stochastic resonance in a single-photon emitter}
\author{H. Mannel}
\email{hendrik.mannel@uni-due.de}

\author{J. Zöllner}
\affiliation{Faculty of Physics and CENIDE, University of Duisburg-Essen, 47057 Duisburg, Germany}
\author{E. Kleinherbers}
\affiliation{Department of Physics and Astronomy, University of California, Los Angeles, California 90095, USA}
\author{M. Zöllner}
\affiliation{Faculty of Physics and CENIDE, University of Duisburg-Essen, 47057 Duisburg, Germany}

\author{N. Schwarz}
\affiliation{Faculty of Physics and CENIDE, University of Duisburg-Essen, 47057 Duisburg, Germany}

\author{F. Rimek}
\affiliation{Faculty of Physics and CENIDE, University of Duisburg-Essen, 47057 Duisburg, Germany}

\author{A. D. Wieck}

\author{A. Ludwig}
\affiliation{Lehrstuhl für Angewandte Festk\"orperphysik, Ruhr-Universit\"at Bochum, 44780 Bochum, Germany}

\author{A. Lorke}

\author{J. König}

\author{M. Geller}
\affiliation{Faculty of Physics and CENIDE, University of Duisburg-Essen, 47057 Duisburg, Germany}

\date{\today}

\maketitle

\newpage

\section{Sample structure}

\begin{figure}
    \centering
    \includegraphics[width=0.75\linewidth]{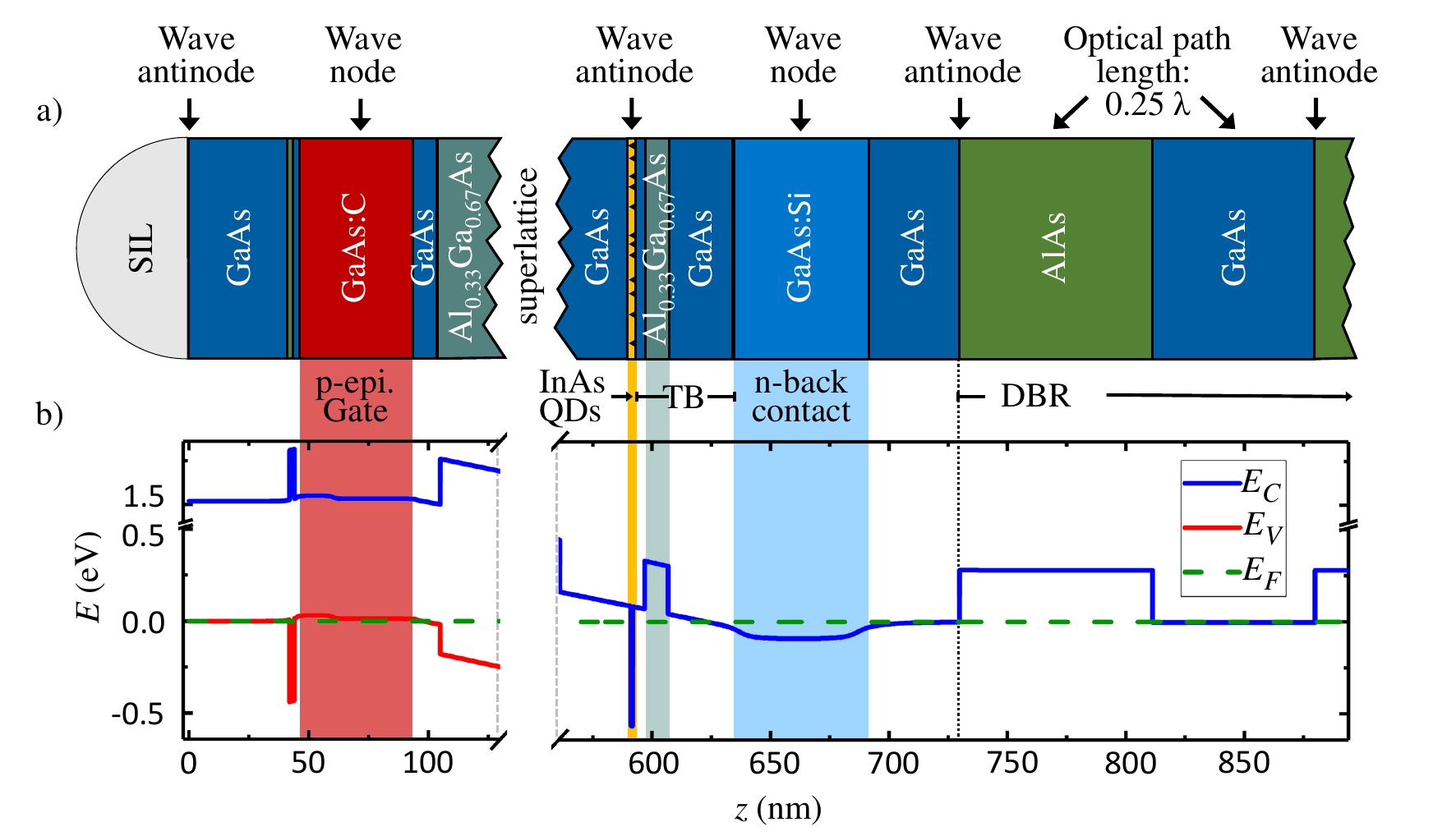}
    \caption{\label{fig:enter-label} {\bf Sample structure and the calculated band structure.} {\bf a} Starting from the bottom on the right-hand side, the diode structure consist of a DBR, a back contact (GaAs doped with Si), a gate (GaAs doped with C), quantum dots (QDs), and an AlGaAs/GaAs superlattice to limit current flow through the sample.  {\bf b} The conduction band edge is represented with a blue line, while the valence band edge is shown in red. The Fermi energy is indicated by a dashed green line. Adapted from Lochner et al.\cite{Lochner.2019}.}
\end{figure}

Figure \ref{fig:enter-label} schematically illustrates the sample structure in {\bf a} from the bottom (right side) to the top (left side) with the corresponding calculated band structure (1D Poisson solver).  The conduction $E_C$ and valence $E_V$ band edge and the Fermi energy $E_F$ are shown in Fig.~\ref{fig:enter-label}{\bf b}. Fig.~\ref{fig:enter-label} is adapted from Lochner et al.~\cite{Lochner.2019}. Starting from the bottom on the left side, the main components of the sample structure include a distributed Bragg reflector (DBR), an electron reservoir, a tunneling barrier, quantum dots (QDs), a superlattice, and the top gate. The DBR consists of 16 alternating layers of GaAs and AlAs to enhance coupling between the QDs and the laser, thereby increasing photon collection efficiency. The back contact comprises a $50\,$nm thick, heavily Si-doped GaAs layer (light blue at position around $z=650\,$nm), which serves as both a 3D electron reservoir and the back contact. The QDs are weakly coupled to this reservoir via a tunneling barrier, consisting of $30\,$nm of GaAs, $10\,$nm of AlGaAs, and $5\,$nm of GaAs (at $z=600\,$nm). Above the QDs, a superlattice separates them from a $45\,$nm C-doped GaAs layer, which functions as the top gate. The AlAs/GaAs short period superlattice serves primarily as a current-blocking barrier. Moreover, the lattice structure suppresses upward segregation of impurities and improves the layer smoothness. After growth and processing, a solid immersion lens (SIL) is positioned on top of the structure.

\section{Telegraph stream with photon histogram}

\begin{figure}
	\centering
	\includegraphics{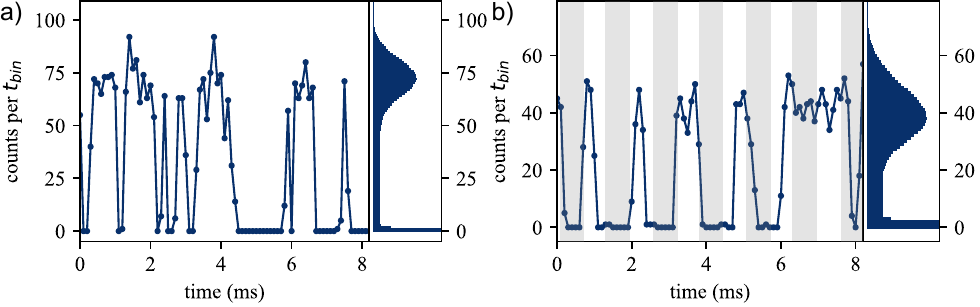}
	\caption{\label{FigStream}{\bf Random telegraph signals.} \textbf{a} A telegraph stream without external modulation. The histogram on the right shows the photon statistics for the full stream of 15 minutes. \textbf{b} Telegraph stream with $16\,$mV modulation amplitude and $796\,$Hz modulation frequency $f_{mod}$. The gray bars indicate a high $V_g$.}
\end{figure}

Figure \ref{FigStream} shows two telegraph streams using a binning time of $\Delta t_{bin}=100\,\mu$s for an unmodulated (in {\bf a}) and modulated electron tunneling dynamics (with modulation frequency of $796\,$Hz and $16\,$mV amplitude) in {\bf b}. The grey-shaded areas correspond to time intervals where the applied gate voltage, denoted by $V_g$, of the modulation is at a high voltage. This demonstrates the correlation between voltage modulation and charge-state switching. The complete telegraph streams have a total length of 15 min. The histograms on the right-hand side of Figs.~\ref{FigStream}{\bf a} and {\bf b} have the same scaling and are used to set the threshold counts to distinguish between charged (low photon counts) and uncharged QD (higher photon counts). The histograms do not contain information about the regularity of the switching.  A more detailed analysis of the evaluation process of unmodulated telegraph signals can be found in Kerski et al.~\cite{Kerski.2023}.

\section{Gate voltages of the modulation}

All modulation was performed with a square function with a duty cycle of 50\% for the different frequencies. The exact gate voltages as well as the specific tunneling rates for the theoretical model are summarized in Table \ref{tab:gatevoltages}.

\begin{table}[h]
    \centering
    \begin{tabular}{c | c c c c c}
        Modulation amplitude (mV) & 1 & 6 & 11 & 16 & 87 \\
        \hline
        Lower gate voltage (mV) & 493.5 & 491 & 488.5 & 486 & 447 \\
        \hline
        $\gamma_{In,low}$ (kHz) & 2.00& 1.41&0.97 & 0.68 & 0.37$^{*}$ \\
        \hline
        $\gamma_{Out,low}$ (kHz) & 2.07& 2.31&2.61 & 2.52 & 3.83$^{*}$ \\
        \hline
        Upper gate voltage (mV) & 494.5 & 497 & 499.5 & 502 & 534 \\
        \hline
        $\gamma_{In,high}$ (kHz) & 2.24 &2.71& 3.80 & 4.52 & 5.78$^{*}$ \\
        \hline
        $\gamma_{Out,high}$ (kHz) & 1.89&1.35& 1.04 & 0.84 & 0.07$^{\dagger}$ \\
        \hline
        Average gate voltage (mV) & 494 & 494 & 494 & 494  & 490.5\\
        \hline
        Resonance frequency (Hz) & - & 796 & 796 & 796 & 378
    \end{tabular}
    \caption{Overview of the gate voltages chosen for the modulation and the corresponding tunneling rates used for the theoretical model.}
    \label{tab:gatevoltages}
\end{table}



The tunneling rates for $1$-$16\,$mV were determined via waiting time distributions from telegraph signals measured at constant gate voltages. The tunneling rates for $87\,$mV marked by $(*)$ were measured via pulsed measurement scheme presented in Zollner et al.\cite{Zollner.2024}. The tunneling rate $\gamma_{out,high}$ for $87\,$mV marked by $(\dagger)$ was calculated using the fit from Fig. 2. This is due to the fact that the measurement via the pulsed measurement scheme becomes unreliable for very large gate voltages.



\section{Probability distribution of electron tunneling}

\begin{figure}
	\centering
	\includegraphics{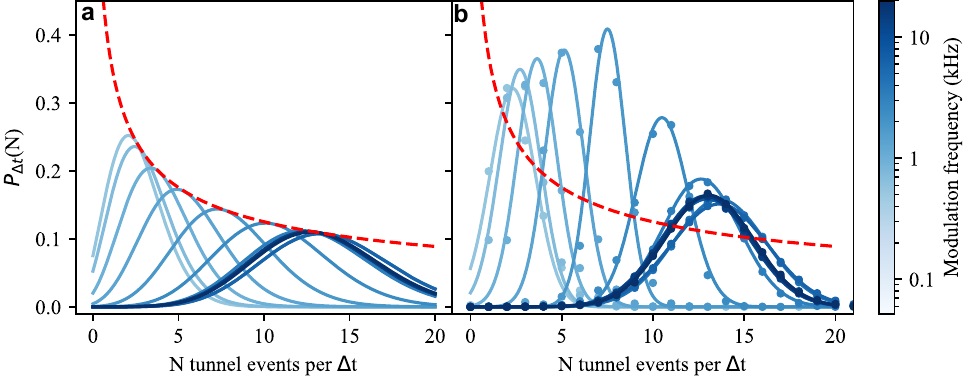}
	 \caption{\label{FigProb} {\bf Poisson distributions in comparison with the quantum stochastic resonance.} {\bf a} Poisson distributions corresponding to the same mean values as the experimental data for increasing modulation frequency. The red dashed line indicates the maximum of each individual Poission distribution. {\bf b} The measured distribution as dots with a Gaussian fit as guide to the eye (since the Poisson distribution is inadequate). As the distribution area is normalized, any distribution above the red line corresponds to a Fano factor below 1.}
\end{figure}

A Poisson distribution is defined as $P_{N}^{Poisson}= \lambda^N e^{-\lambda}/N!$,
where $N$ is the number of events and $\lambda$ is the expectation value of tunneling-out events in a time interval of length $\Delta t$. 
In Fig.~\ref{FigProb}a, Poisson distributions for different $\lambda$ are shown with a red dashed guide to the eye for the maximum value. In Fig.~\ref{FigProb}b, the probability distributions of the electron tunneling events of the $87\,$mV modulation are shown as dots. The blue lines are fits of a Gaussian distribution to the data and the red dashed line is the same guide to the eye as shown in Fig.~\ref{FigProb}a for comparison to the Poisson distribution. Since the area of the distribution is normalized to unity, a smaller Fano factor results in a higher maximum value. At a modulation frequency of $f_{mod}=796$ Hz, the transport becomes more regular, i.e. the distribution becomes narrower and shows here the quantum stochastic resonance.

\section{Factorial cumulants in terms of moments}

The combination of Eq.~(2) with (3) of the main paper yields the expression 
\begin{align}
    C_{F,m}= \left[ \partial ^m_z \ln \left( \sum_N z^N P_N \right) \right]_{z=1}
    \label{eq:FakCu}
\end{align}
for the factorial cumulants.
It is straightforward to express them in terms of the moments $\langle N^m \rangle = \sum_N N^m P_N$.
We find
\begin{subequations}
\begin{align}
    C_{F,1} = & \langle N \rangle
    \\
    C_{F,2} = & \langle N^2 \rangle - \langle N \rangle^2 - \langle N \rangle
    \\
    C_{F,3} = & \langle N^3 \rangle - 3\langle N^2 \rangle \langle N \rangle + 2 \langle N \rangle^3 - 3 \langle N^2 \rangle + 3 \langle N \rangle^2 + 2 \langle N \rangle
    \\
    C_{F,4} = & \langle N^4 \rangle - 4\langle N^3 \rangle \langle N \rangle - 3 \langle N^2 \rangle^2 + 12 \langle N^2 \rangle \langle N \rangle^2 - 6 \langle N \rangle^4 
    \nonumber \\ & -6 \langle N^3 \rangle + 18 \langle N^2 \rangle \langle N \rangle - 12 \langle N \rangle^3 + 11 \langle N^2 \rangle - 11 \langle N \rangle^2- 6\langle N \rangle
\end{align}
\end{subequations}
for the first four factorial cumulants.

\section{Factorial cumulants for special distributions}

\subsection{Poisson distribution}
The Poisson distribution
\begin{align}
    P_N^{Poisson} = \frac{\lambda^N}{N!} e^{-\lambda} 
\end{align}
is parametrized by one number, namely $\lambda$.
The corresponding generating function is
\begin{align}
    S_F^{Poisson}(z) = \ln \left(\sum_N z^N \frac{\lambda^N}{N!} e^{-\lambda}\right) = \ln e^{\lambda (z-1)} = \lambda (z-1) 
\end{align}
which implies
\begin{align}
    C_{F,1}^{Poisson}= \lambda \qquad \text{and} \qquad C_{F,m}^{Poisson}=0 \quad \text{for} \quad m\ge 2 \, .
\end{align}
This shows that the parameter $\lambda$ is nothing but the mean value and that higher-order factorial cumulants of a given distribution measure deviations from a Poisson distribution.

We note here that the condition $C_{F,2}=0$, which is equivalent to $F=1$, is necessary but not sufficient for a Poisson distribution.
This is illustrated by the example $P_N= \frac{1}{2} \left( \delta_{N,0} + \delta_{N,2} \right)$, which yields $S_F(z) = \ln (1+z^2) - \ln 2$ and thus $C_{F,1}=1$ and $C_{F,2}=0$ but $C_{F,3}= -1 \neq 0$.

\subsection{Kronecker distribution}

Next, we consider a completely regular distribution with the same mean value $\lambda$, which is given by the Kronecker distribution 
\begin{align}
    P_N^\delta = \delta_{N,\lambda} 
\end{align}
where we assume $\lambda$ to be integer.
The corresponding generating function reads
\begin{align}
    S_F^\delta(z) = \ln \left(\sum_N z^N \delta_{N,\lambda}\right) = \lambda \ln z 
\end{align}
which implies
\begin{align}
    \qquad C_{F,m}^\delta=(-1)^{m-1} (m-1)! \lambda \, .
\end{align}

We note that the condition $C_{F,2}=-C_{F,1}$ is both necessary and sufficient for a Kronecker distribution, because this case corresponds to a vanishing Fano factor, $F = \langle \left( N - \langle N \rangle \right)^2 \rangle =0$, and therefore vanishing variance.

\section{Details of the theoretical model}

The charge dynamics of the quantum dot is described within a two-state model.
The label $\chi=0,1$ for the two states describes the quantum-dot charge.
For the full counting statistics, we are interested in the probability distribution $P_N$ that $N$ tunneling-out events have taken place within a given time interval.
In a first step, we split this distribution into the state-resolved distributions $P_{\chi,N}$ with $P_N=P_{0,N}+P_{1,N}$.
Their time evolution is governed by the master equations
\begin{align}
\begin{split}
	\dot P_{0,N}(t) &= -\gamma_{In}(t)P_{0,N}(t)+\gamma_{Out}(t)P_{1,N-1}(t)\\
	\dot P_{1,N}(t) &= \gamma_{In}(t)P_{0,N}(t)-\gamma_{Out}(t)P_{1,N}(t)
\end{split}
\label{eq:master_N}
\end{align}
which are coupled in $N$ since tunneling-out of an electron increases the counting by one.
Decoupling is achieved by making use of the $z$-transforms $P_{\chi,z}=\sum_N z^N P_{\chi,N}$.
Multiplying Eq.~(\ref{eq:master_N}) with $z^N$ and summing over $N$ leads to the differential equation
\begin{align}
    \frac{d}{dt}
\begin{pmatrix}P_{0,z}(t)\\P_{1,z}(t)\end{pmatrix}
    =\mathbf{W}_z(t) 
    \begin{pmatrix}P_{0,z}(t)\\P_{1,z}(t)\end{pmatrix},
\label{eq:master_z}
\end{align}
which is decoupled in $z$ and governed by the transition matrix
\begin{align}
    \mathbf{W}_z(t)=
    \begin{pmatrix}
        -\gamma_{In}(t)&&z \gamma_{Out}(t)\\
        \gamma_{In}(t)&&-\gamma_{Out}(t)
    \end{pmatrix}
\label{eq:W_z}
\end{align}
that contains the time-dependent tunneling rates.
The counting variable $z$ only appears in the upper right matrix element.

The formal solution of Eq.~(\ref{eq:master_z}) with initial time $t_0$ is given by
\begin{align}
    \begin{pmatrix}P_{0,z}(t)\\P_{1,z}(t)\end{pmatrix}
    =\mathbf{\Pi}_z(t,t_0)
    \begin{pmatrix}P_{0,z}(t_0)\\P_{1,z}(t_0)\end{pmatrix},
\end{align}
where we introduced the propagator
\begin{align}
    \mathbf{\Pi}_z(t,t_0)
    =\mathcal{T}\exp \left(\int_{t_0}^{t}\mathbf{W}_z(t')dt'\right)
    \label{eq:propagator}
\end{align}
from the initial time $t_0$ to time $t$ by making use of the time-ordering operator $\mathcal{T}$.
Choosing the initial time $t_0$ as the time at which counting starts implies that $P_{\chi,N}(t_0)=P_\chi(t_0) \delta_{N,0}$.
This means that the initial values $P_{\chi,z}(t_0)=P_\chi(t_0)$ are independent of $z$ and given by the probability to find the quantum dot in charge state $\chi$. 

For an arbitrary time dependence of the tunneling rates, the time ordering complicates the numerical evaluation of the propagator.
In our case, however, the gate voltage is modulated with a square function.
Therefore, the tunneling rates $\gamma_{In}(t)$ and $\gamma_{Out}(t)$ are piecewise constant in the intervals $[0;\frac{T}{2}]$, $[\frac{T}{2},T]$, $\ldots$, which suggests to divide for $0\le t_0 < \frac{T}{2}$ the full propagator into the product 
\begin{align}
    \mathbf{\Pi}_z(t,t_0)=\mathbf{\Pi}_z\left( t,\frac{MT}{2}\right) ~\cdots~ \mathbf{\Pi}_z\left(\frac{3T}{2},T\right)\mathbf{\Pi}_z\left(T,\frac{T}{2}\right)\mathbf{\Pi}_z\left(\frac{T}{2},t_0\right),
\end{align}
in which each factor is easily calculated by performing a matrix exponentiation, where $M$ is the smallest number that fulfills
$t \le \frac{M+1}{2}T$.
In case of $\frac{T}{2}\le t_0 < T$, the two rightmost factors have to be replaced by the single factor $\mathbf{\Pi}_z\left(T,t_0\right)$.

The procedure described so far assumes infinite time resolution of the detector, i.e., every single tunneling-out event is counted by introducing a factor $z$ in the propagator.
In the experiment, however, the time resolution is limited by the chosen binning time $t_{bin}$ of the photon counting.
As a result, some of the real charge-switching events are missed.
To model this effect, we follow the procedure described in Kleinherbers et al.~\cite{Kleinherbers.2022,Kleinherbers.2023}.
We first discretize the time evolution by dividing the full propagator into $(t-t_0)/t_{bin}$ propagators of length $t_{bin}$ each,
\begin{align}
    \mathbf{\Pi}_z(t,t_0)=\mathbf{\Pi}_z\left( t, t-t_{bin}\right) ~\cdots~ \mathbf{\Pi}_z\left(t_0+t_{bin},t_0\right) \, .
\end{align}
Then, we replace each of the short propagators $\mathbf{\Pi}_z(t'+t_{bin},t')$ by
\begin{align}
    \mathbf{\Pi}^{miss}_z(t'+t_{bin},t'):=
    \begin{pmatrix}
        \left[\mathbf{\Pi}_{z=1}(t'+t_{bin},t')\right]_{00}&&z\left[\mathbf{\Pi}_{z=1}(t'+t_{bin},t')\right]_{01}\\
        \left[\mathbf{\Pi}_{z=1}(t'+t_{bin},t')\right]_{10}&&\left[\mathbf{\Pi}_{z=1}(t'+t_{bin},t')\right]_{11}
    \end{pmatrix} \, .
\end{align}
It is obtained from $\mathbf{\Pi}_z(t'+t_{bin},t')$ by switching off the counting \textit{during} the time propagation through setting $z=1$ and only checking \textit{after} the propagation whether the charge state has switched from a charged to an uncharged quantum dot through multiplying $z$ to the upper right matrix element.
While the continuous propagator $\mathbf{\Pi}_z(t'+t_{bin},t')$ contains contributions of high powers in $z$ (which corresponds to a large number of counts), the discretized propagator $\mathbf{\Pi}^{miss}_z(t'+t_{bin},t')$ is only \textit{linear} in $z$ since at most \textit{one} switching event is counted.

\begin{figure}
    \centering
    \includegraphics[width=0.7\linewidth]{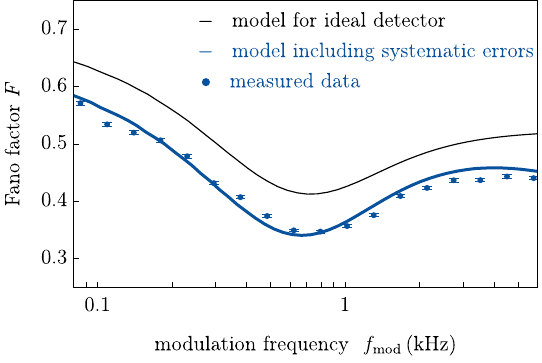}
    \caption{\textbf{Influence of limited time resolution and statistical uncertainties on Fano factor.} Fano~factor as a function of the modulation frequency for a modulation amplitude of $16\,$mV, a time interval $\Delta t=20$ms and a binning time $t_{bin}=100\, \text{$\mu$s}$ (blue data points) compared to theoretical models for the same binning time (blue line) and ideal time resolution $t_{bin}=0\,\text{$\mu$s}$ (black line). The statistical uncertainties caused by the finite length of the photon streams, as described by Eq.~(\ref{eq:uncertainties}), are depicted by error bars.
    \label{fig:fano_t_bin}}
\end{figure}

With this propagator, we evaluate $P_{0,z}(t)$ and $P_{1,z}(t)$.
To obtain the generating function for the factorial cumulants for a time interval of length $\Delta t$, we set $t=t_0+\Delta t$, average the probability distribution for fixed $\Delta t$ over the initial time $t_0$, and finally use
\begin{align}
    S_F(z) = \ln \left( P_{0,z}+ P_{1,z} \right) \, .
\end{align}
As an illustration of the effects of limited time resolution, Fig. \ref{fig:fano_t_bin} shows the predicted Fano factor for an ideal detector $t_{bin}=0\,\mu\text{s}$ and the experimental time resolution $t_{bin}=100\,\mu\text{s}$.

\section{Statistical uncertainties of factorial cumulants}
\begin{figure}
    \centering
    \includegraphics[width=0.7\linewidth]{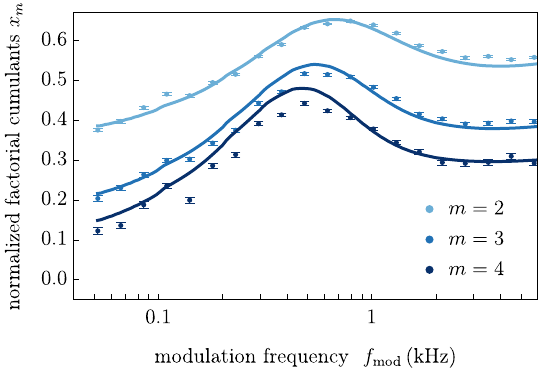}
    \caption{\textbf{Statistical uncertainties of the factorial cumulants.}
    The second, third and fourth normalized factorial cumulant $x_m$ as a function of the modulation frequency for the experimental data as blue dots and the model as solid blue lines at a modulation
    amplitude of $16\,$mV and a time interval length of $\Delta t = 10\,$ms. The statistical uncertainties caused by the finite length of the photon streams, as described by Eq.~(\ref{eq:uncertainties}), are depicted by error bars.
    \label{fig:fig_cumulants_stat}}
\end{figure}

The finite length of each photon stream $t_{stream}$ results in statistical uncertainties in all resulting factorial cumulants and the Fano factor. For large time intervals, these can be estimated by \cite{Kleinherbers.2022,Kleinherbers.2023}

\begin{align}
    \Delta C_{F,m}(\Delta t)= \sqrt{\frac{m!\Delta t}{t_{stream}}} \left(\left\langle N^2\right\rangle-\left\langle N\right\rangle^2\right)^{\frac{m}{2}}.
    \label{eq:uncertainties}
\end{align}
Figures \ref{fig:fano_t_bin} and \ref{fig:fig_cumulants_stat} show the statistical uncertainties of the Fano factor, as well as the second, third and fourth order factorial cumulants.

\section{Ordinary cumulants}

\textit{Factorial} cumulants measure deviations from a Poisson distribution and are, thus, suitable for \textit{discrete} stochastic variables.
In case of a \textit{continuous} stochastic variable, it is more natural to use \textit{ordinary} cumulants, since they measure deviations from a Gauss distribution.
Ordinary cumulants are derived from
\begin{align}
    C_m= \left[ \partial ^m_\chi S(\chi)\right]_{\chi=0} 
\end{align}
with the generating function
\begin{align}
    S(\chi) = \ln \left( \sum_N e^{N\chi} P_N \right) \, .
\end{align}
The expression for the ordinary cumulants is obtained from the corresponding expression Eq.~(\ref{eq:FakCu}) for factorial culumants by replacing $z$ with $e^{N\chi}$, performing the derivative with respect to $\chi$ and setting $\chi=0$ at the end.

In terms of moments, the first four ordinary cumulants are given by
\begin{subequations}
\begin{align}
    C_1 = & \langle N \rangle
    \\
    C_2 = & \langle N^2 \rangle - \langle N \rangle^2
    \\
    C_3 = & \langle N^3 \rangle - 3\langle N^2 \rangle \langle N \rangle + 2 \langle N \rangle^3
    \\
    C_4 = & \langle N^4 \rangle - 4\langle N^3 \rangle \langle N \rangle - 3 \langle N^2 \rangle^2 + 12 \langle N^2 \rangle \langle N \rangle^2 - 6 \langle N \rangle^4 
    \, .
\end{align}
\end{subequations}

The generating function for a Poisson distribution with mean value $\lambda$ is given by the expression
$S^{Poisson}(\chi )= \lambda \left( e^\chi -1 \right)$.
As a result, all ordinary cumulants acquire the same value $C_m^{Poisson}=\lambda$.
In contrast, the Kronecker distribution is characterized by $S^\delta(\chi )= \lambda \chi $.
This implies $C_1^\delta= \lambda$ as well as $C_m^\delta=0$ for $m \ge 2$.

One may be tempted to use, in analogy to $x_m$, the \textit{normalized ordinary cumulants}
\begin{align}
    y_m := \frac{C_m}{C_1}
\end{align}
as a measure of how regular regular the switching is, where $y_m=1$ (instead of $x_m = 0$ for $m \ge 2$) indicates purely random and $y_m=0$ for $m \ge 2$ (instead of $x_m= 1$) purely regular switching.

\begin{figure}
    \centering
    \includegraphics[width=0.7\linewidth]{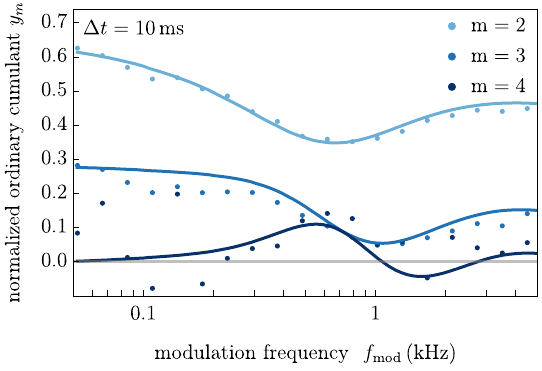}
    \caption{\textbf{Normalized ordinary cumulants.}
    The normalized second-, third- and fourth-order ordinary cumulants $y_m$
    as a function of the modulation frequency for a modulation amplitude of $16\,$mV and a time interval length of $\Delta t=10$ms. The experimental data are displayed as blue dots and the model as solid blue lines. 
    Although completely regular switching implies $y_m=0$ for all $m\ge 2$, it is not possible to identify quantum stochastic resonance through minima of $|y_m|$. 
    \label{fig:ordinary_cumulants}}
\end{figure}

The problem with that approach is that ordinary cumulants universally show oscillations as a function of any system parameter~\cite{Flindt.2009}, which naturally also includes the modulation frequency $f_{mod}$.
This is illustrated in Fig.~\ref{fig:ordinary_cumulants}, which shows the second, third and fourth normalized ordinary cumulant $y_m$ as a function of the modulation frequency $f_{mod}$ for a modulation amplitude of $16\,$mV. 
The experimental data is represented by blue dots while the theoretical model is given by solid, blue lines. 
We find stronger deviations between experiment and theory than we found for factorial cumulants.
This reflects the fact that ordinary cumulants are more sensitive to detector imperfections than factorial cumulants~\cite{Kleinherbers.2022,Kleinherbers.2023}.
More important, however, are the disturbing oscillations.
The fourth ordinary cumulant crosses the zero line twice, but both times at a frequency that is quite far away from the resonance frequency for the Fano factor.
This renders higher-order ordinary cumulants as indicators of quantum stochastic resonance useless.

\clearpage

\bibliographystyle{naturemag}
\bibliography{StochRes-PaperBib-Paul}